\documentclass[preprint]{article}

\usepackage{blindtext} 

\usepackage{graphicx}
\usepackage{subfigure}

\usepackage[sc]{mathpazo} 
\usepackage[T1]{fontenc} 
\usepackage{microtype} 

\usepackage[english]{babel} 

\usepackage[hmarginratio=1:1,top=32mm,columnsep=20pt]{geometry} 
\usepackage[hang, small,labelfont=bf,up,textfont=it,up]{caption} 
\usepackage{booktabs} 

\usepackage{lettrine} 

\usepackage{enumitem} 
\setlist[itemize]{noitemsep} 

\usepackage{abstract} 

\usepackage{titlesec} 
\renewcommand\thesection{\Roman{section}} 
\renewcommand\thesubsection{\roman{subsection}} 
\titleformat{\section}[block]{\large\scshape\centering}{\thesection.}{1em}{} 
\titleformat{\subsection}[block]{\large}{\thesubsection.}{1em}{} 

\usepackage{fancyhdr} 
\usepackage{titling} 

\usepackage{hyperref} 

\usepackage{graphicx}

\usepackage{hyperref} 

\newcommand{\beq}{\begin{equation}}
\newcommand{\eeq}{\end{equation}}

\newcommand{\ben}{\begin{enumerate}}
\newcommand{\een}{\end{enumerate}}

\pretitle{\begin{center}\huge\bfseries} 
\posttitle{\end{center}} 
\title{Plea to publish less} 
\author{%
\textsc{Navinder Singh}\thanks{Cell Phone/WhatsApp: +919662680605} \\[1ex] 
\normalsize Physical Research Laboratory, Ahmedabad, India. \\ 
\normalsize \href{mailto:navinder.phy@gmail.com}{navinder.phy@gmail.com} 
}
\date{\today} 
\renewcommand{\maketitlehookd}{%
\begin{abstract}
Recent research\cite{fort} has shown that in the 20th century there is an exponential growth of the number of published scientific papers, but new ideas has only a linear growth with time. The 19th and the first half of the 20th century saw major scientific discoveries. But from the second half of the 20th century, the number of publications far exceeds the number of impactful discoveries.  Does an exponentially growing number of publications indicate an element of pathological research? Pressure to publish a large number of papers has led to the phenomena of overproduction, unnecessary fragmentations, overselling, predatory journals (pay and publish), clever plagiarism, and deliberate obfuscation of scientific results so as to sell and oversell. This is harming the healthy scientific culture of passion driven research. It is an urgent problem which needs attention of the whole of the scientific community. Some ways to mitigate these grave problems are discussed.
\end{abstract}
}

\begin{document}
\maketitle


\vspace{2pc}
\noindent{\it Keywords}: The changing culture of the practice of science; floods of papers; limited number of impactful discoveries; race for overproduction; unnecessary fragmentation; overselling; predatory journals; predatory conferences; clever plagiarism; re-discovering and re-casting; overloaded editorial boards.

\section{Changing culture of the practice of science:  before 20th century.}

Science before the turn of the 20th century was practiced by individual researches mostly motivated by the urge to find some "transcendental  truths" in natural phenomena.  Some of the leading discovers made in the 16th, 17th, 18th, and the 19th century are: Overthrowing of the geocentric model of the universe and the introduction of the heliocentric model by Nicolaus Copernicus (1543);  law of inertia (1613), observational astronomy, and the scientific method of experimentation by Galileo Galilei; Kepler's three laws of the planetary motion (1619);  Many pioneering discoveries in almost all the fields of natural science and mathematics by Issac Newton (Principia 1687);  atomic theory in chemistry by John Dalton (1805); magnetic effects of electric current by Oersted (1820). Sadi Carnot (1824),  Lord Kelvin (1848), Rudolf Clausius (1850) and  others laid the foundations of classical thermodynamics; Statistical mechanics was developed by Maxwell, Boltzmann, and Gibbs; Investigations of Faraday, Maxwell, and many other investigators lead to the theory of electromagnetism (1862);  X-rays were discovered by Wilhelm Roentgen (in 1895);  Radioactivity by Henri Becquerel (in 1896); Electron by J. J. Thomson (in 1897); etc.

However, data shows that only a few thousand of papers were published\cite{gian}. Majority of the papers published had some impact. And the phenomenon of the collaborative research was very rare (most of the time, papers published were single-authored). 

Majority of professors/investigators use of publish a couple of dozens of papers in their entire scientific career. Consider the example of Josiah Gibbs (1839-1903) who published less than two dozen papers in his entire career. Gibbs case is a typical case of a scientist of that era (in terms of the number of publications).

\section{Physics in 20th century}

From 1900  to 1950, some of the main discoveries made are: Energy quantum by M.  Planck (1900);  nuclear atom by Rutherford (1911); superconductivity by Onnes (1911); special and general relativity by Einstein (1905, and 1915 respectively);  crystal structures and X-ray diffraction (William and Lawrence Bragg, 1915); Quantum mechanics (M. Planck, N. Bohr, A. Einstein, W. Heisenberg, E. Schroedinger, M Born, P. Dirac and others); Raman effect (by C. V. Raman); Dirac's theory of electrons and positrons; nuclear fission and fusion; Quantum electrodynamics; etc. The later half of the 19th century and the first half of the 20th century can be said "the golden age" of physics.

From 1950  to Current times, some of the main discoveries made are : Nuclear and particle physics; solid state physics; semiconductors and transistors;  electroweak theory and the emergence of standard model; Bardeen-Cooper-Schrieffer (BCS) theory of superconductivity; parity violation; quantum Hall effects; renormalization; QCD; BEC realized; Higgs boson observed; gravitational waves observed, etc. 

After WWII, there is an outpouring of the scientific activity. This is due to technological developments, new instruments of investigations, and due to the vast applicability of quantum mechanics into atomic, solid state, and nuclear physics.  This outpouring is evident from how the scientific journals evolved from pre-WWII era to post-WWII era. Take for example, the Phys. Rev.  The Phys Rev started on 1893, as a single volume. In 1958, PRL was introduced; and in 1970, PRA, PRB, PRC, and PRD were introduced.  Currently APS is publishing an array of journals: Phys. Rev. A, B, C, D. Phys. Rev. E; Physics; Phys. Rev. X; Phys. Rev. Applied; Phys. Rev. Fluids; Phys. Rev. Materials; Phys. Rev. Research; PRX Quantum etc.

In the year 1910 Phys. Rev. published articles with 1484 printed pages in total, whereas in 2019, only the Phys. Rev. B (one journal out of the whole array) published 98,398 articles in just six months (from Jan 2011 to June 2019). If each article contains roughly 10 pages, then this amounts to {\it one million} pages!

\section{What has gone wrong in current times?  (Are we going from facts to pathology?)}

There were some thousands of papers published before 1900. By 1960, two million scientific papers have been published! Current times, over two million scientific papers are being published every year! What was produced between 1650 to 1960 (over three hundred years) is now being produced every year (in terms of volume, not number of discoveries)\cite{gian}. But the number of impactful discoveries made in the last half of the 20th century to the current time is much less than the number of papers published in this time period (in comparison to that in the 19th and the first half of the 20th century). Concrete data\cite{fort} shows that there is a linear growth of ideas but there is an exponential growth of the number of papers published (figure 1).

\begin{figure}[h!]
\begin{center}
\begin{tabular}{cc}
\includegraphics[width =5.5cm]{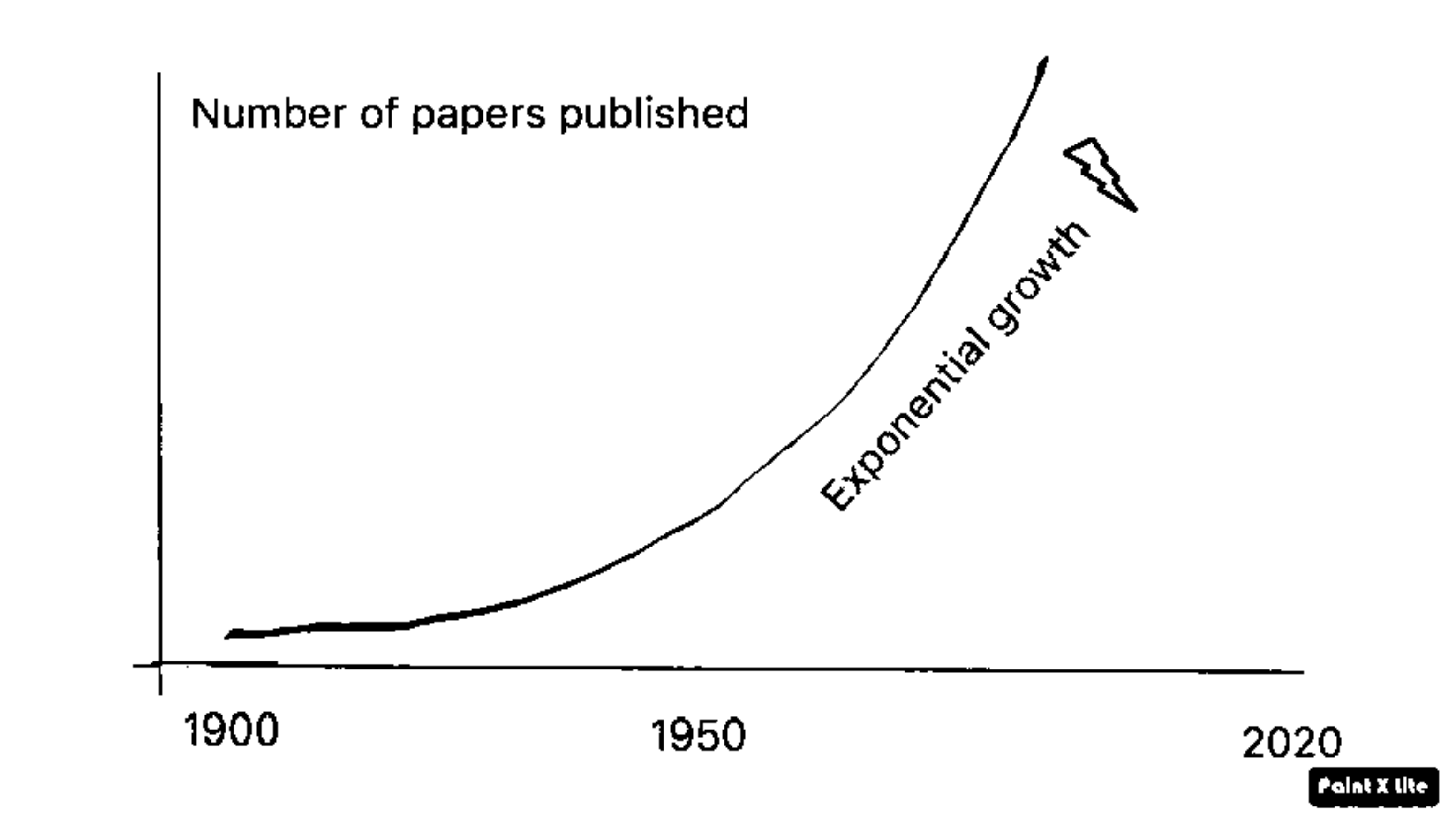}&
\includegraphics[width =5.5cm]{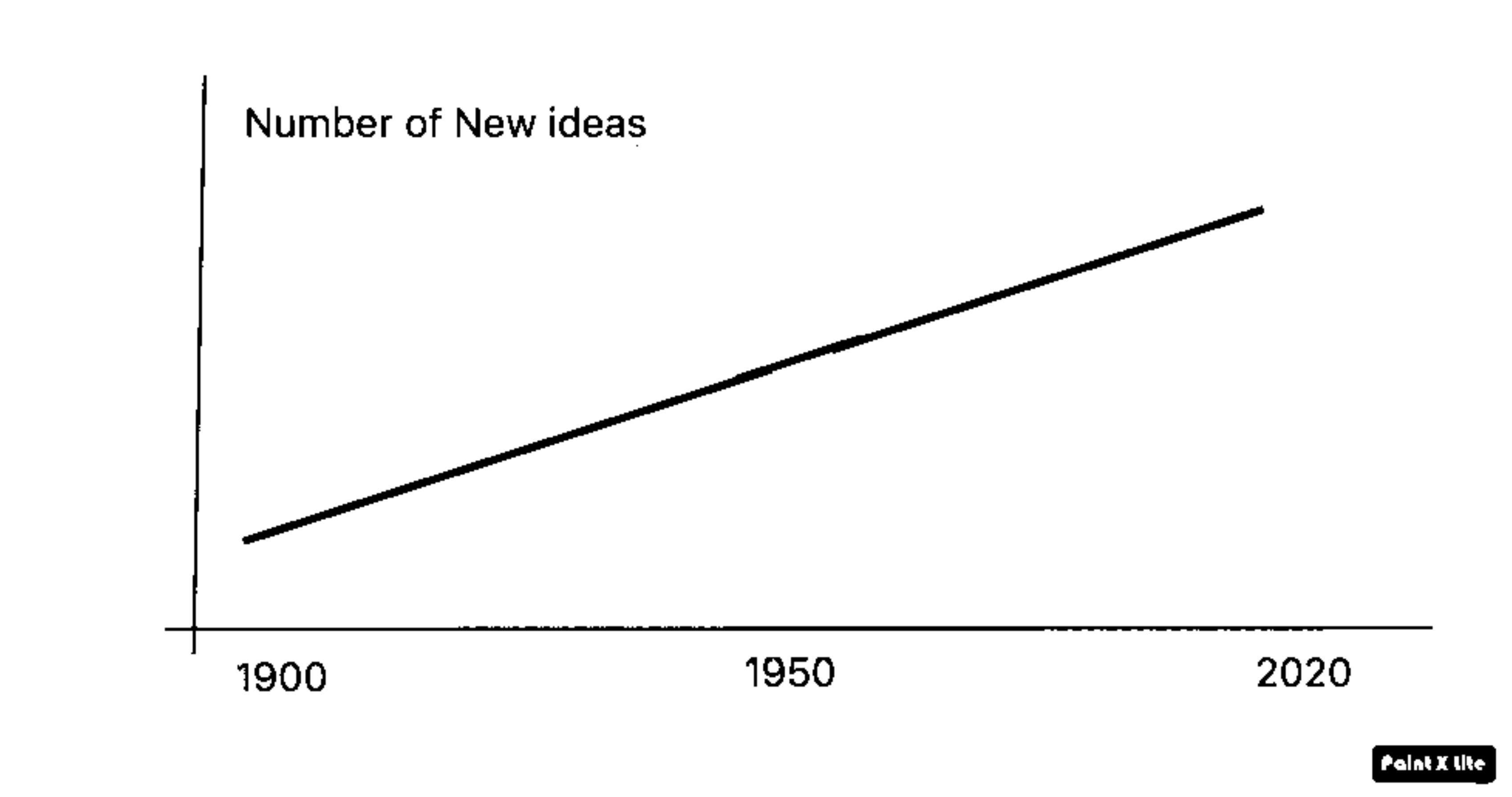}\\
(a)&(b)\\
\end{tabular}
\end{center}
\caption{(a) Exponential growth of paper production.  (b) Only linear growth of ideas. For actual data and graphs, refer to\cite{fort}.}
\label{f1}
\end{figure}

This strange trend in the data raises serious questions: Are we going from facts to pathology? Is it due to scientific developments only (new fields, new subfields)? Or is it due to the ever growing number of active scientists? What is happening?

\section{Reasons (for floods of papers).}

Simplest reason one can give is that there are more number of active scientists working in very competitive environments (as compared to the situation in the 19th and early 20th century). Data shows that in 1930s there was one scientist per 10,000 people in the whole world. This number in 2013 swelled to one scientist per 700 people\cite{gian}.  This is 14 fold increase! There were 3 lakh scientists in 1935. And this number grows to 10 million in 2016. Even if the "production rate" per scientist (that is the average number of papers published per scientist) remains the same, this sheer rise in the number of active scientists could explain the larger production of the number of published papers\footnote{In old culture, majority of professors/investigators use of publish a couple of dozens of papers in their entire career. Currently, majority of professors/investigators publish over 100 papers in their entire scientific career. So average number of papers published {\it per scientist} has also increased, at least in the physical sciences. Collaborations increase the production rate further.}

But the question raised in the previous section: Why are the number of impactful discoveries made much less than the number of papers published?

It could be due to the following reasons:

\ben

\item There are distortions in the practice of science: Old trend: theoretical understanding of the experimental facts (Max Planck and black body radiation; Einstein and photoelectric effect etc). Passion to find some ``transcendental truths''.  New tread: More theoretical ideas and more ``proposed'' experiments. So more papers! More generalizations and more abstractions Imagine a hypothetical situation, do a calculation, produce a paper! This is due to the pressure to publish. This drastically reduce the ratio of the number of impactful discoveries to the total number of papers published.

\item Importance of numbers of papers published in jobs, getting pdf positions, getting funds etc. This put emphasis on numbers along with quality. As we discuss below, the metrics that measure quality are not foolproof.

\item Sheer scientific culture that promotes the number business (refer to footnote 1).

\item Emergence of competitive environments. This further leads to a constant pressure to build huge groups/collaborations etc. This results in overproduction. There is a trend to follow hot topics/fashionable topics/hypes\cite{eva}. Invent strategies to sell and over-sell (intense fights with editors and referees).

\item Old times: Mostly single author or couple of authors; Now majority of papers come out of  collaborations (Collaborations sometimes are required). Collaborations has the potential to increase volume, but whether it leads to an out-put of a great quality publications? Surely a point to ponder over!\cite{henri}.

\item Pressure to publish in reputed journals, rejections, and then the phenomenon of predatory journals: you pay and publish. Easy and low quality publications. Loss of impactful science. And most importantly the misuse of the concept of ``open access".

\item Core open scientific problems are limited in number. There are problems posed by ``Nature" and then there are "man-made" problems. The main scientific problems in theoretical physics can be counted: Beyond standard model physics; What is small CP violation trying to tell us? GUT: ``Wedding" of QCD with electroweak theory; Neutrino mass problem; Cuprate high temperature superconductivity; Physics of heavy fermions metals etc. Electroweak theory is renormalizable, but there is no Quantum theory of gravity which is renormalizable; Can there be a theory using which one can compute the mass of an electron? The problem of dark matter etc.

\begin{figure}[!h]
\begin{center}
\includegraphics[height=5cm]{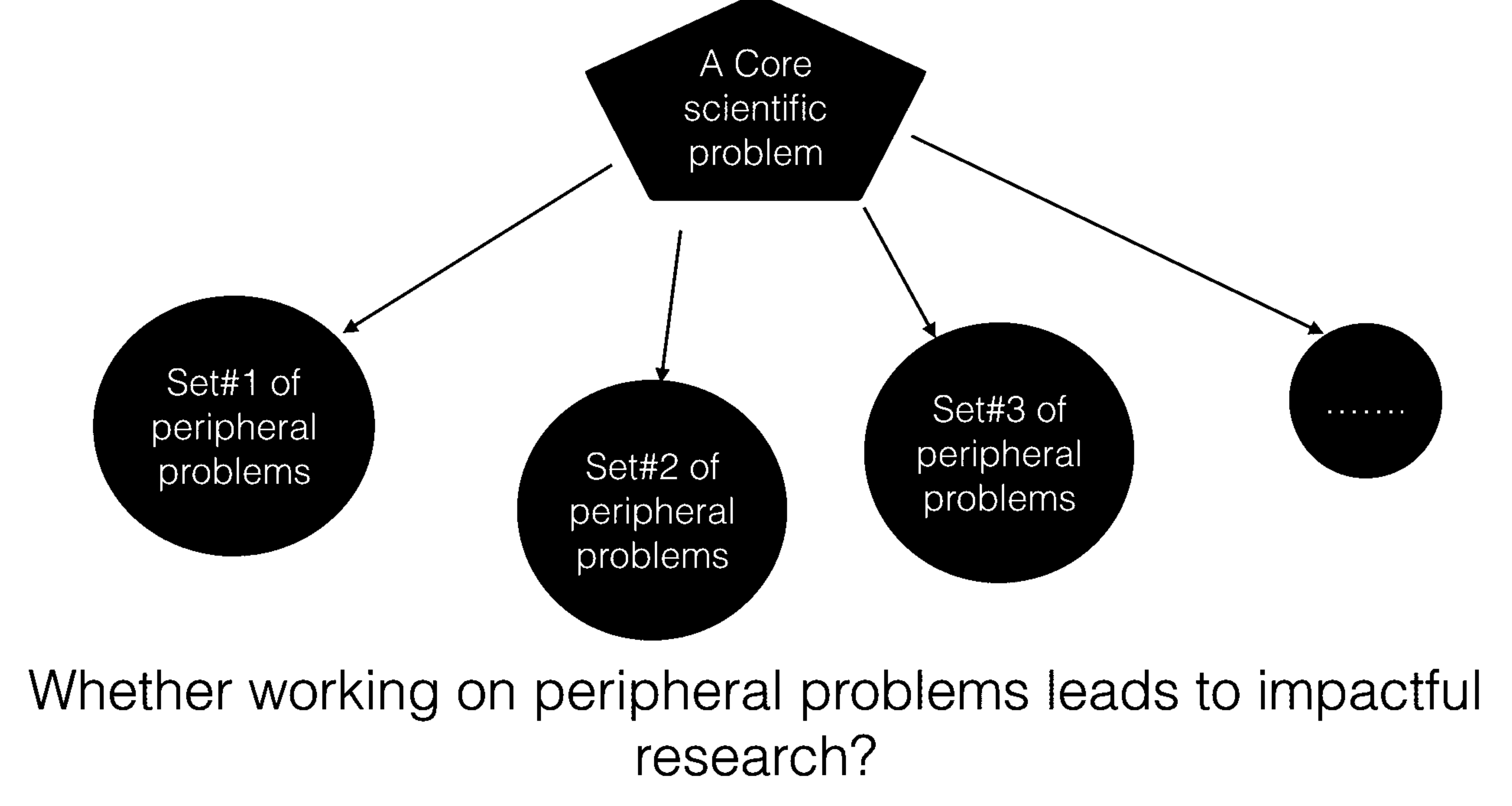}
\caption{Sets of peripheral problems.}
\end{center}
\label{f2}
\end{figure}

To meet the high demands of competition and job issues one is forced to invent easy and peripheral problems (refer to figure (2)).  Working on peripheral problems leads to publications. It can be helpful to a student in his/her career. It can be helpful in promotions/jobs. It may be helpful in getting some awards, fellowship of academies etc. But it leads to a loss of the impactful science. It takes years to solve a hard problem, but an easy problem can be solved in lesser time. The current academic culture supports the latter one, unfortunately!

\item There are experimentally underdetermined problems. When the available input experimental parameters required to validate a theory are much less in number, then there is a large parameter space to play with. Many theoretical models/scenario are possible (as the problem remains observationally/experimentally under-determined). This may lead to burden to literature. Care must be exercised when publishing in such areas.

\item Addiction to publish: I know a person who gets some sort of irritation (mental upsetting) if he does not get a paper in a month's time. 
So every month he must get a paper (low quality, high quality, through students, through collaborators, it does not matter!).
Will that large volume of papers (containing solutions of easy and man-made problems) be useful?
Will this not burden referees/editors/and students who are just entering the field?

\item Vicious cycle:

The more production of papers leads to getting more funding and further leading to bigger collaborations/groups. This in turn leads to a greater production of papers. Thus, a ``vicious cycle" starts! (refer to figure 3). 

\begin{figure}[!h]
\begin{center}
\includegraphics[height=5cm]{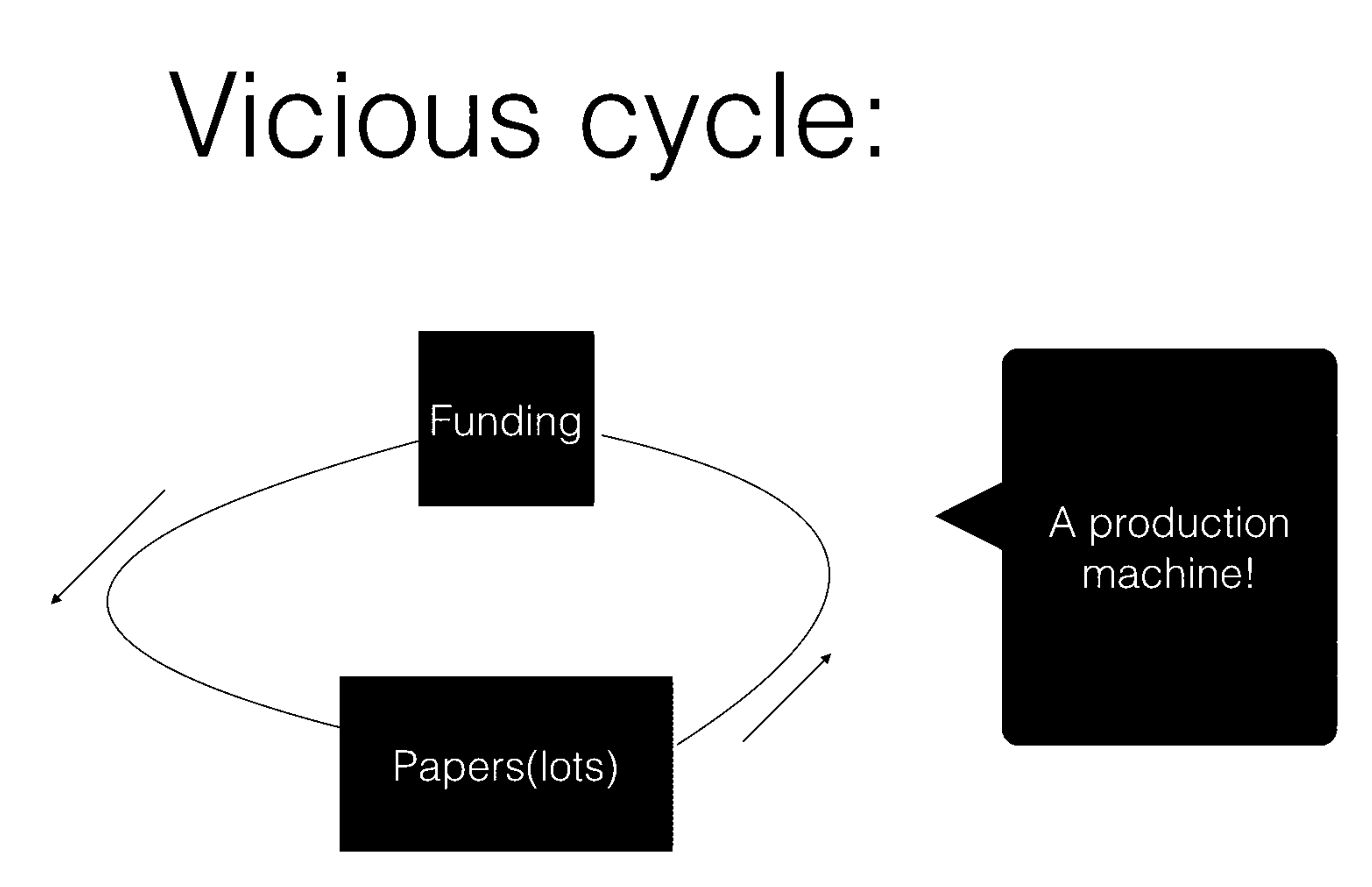}
\caption{A ``vicious cycle".}
\end{center}
\label{f2}
\end{figure}

But there is a difference between an industrial production and scientific solution of a riddle posed by Nature. It is the duty of the funding agencies that they should not put emphasis on the quantity, and while measuring the quality of the scientific output through the standard measures, they must realize that these measures (metrics, impact factors of journals etc) are themselves are distorting the good culture (as discussed below). Actual reading of the scientific papers by an expert panel can help save the situation.

\een


\section{Consequences (on individuals; on sociology; and on science itself).}

Emphasis on the number of publications (or culture of producing a large number of publications) leads to 

\ben

\item a shift of focus from central difficult problems to more peripheral and easy problems.  It is easy to solve an easy problem and more difficult to solve a difficult problem.

\item loss of cohesiveness of a given field and unnecessary fragmentations within the subfields.  Leads to difficulties for a beginner to understand and penetrate into a given field (One has to deal with a huge volume of literature, much of which may not be relevant). The human capacity for information processing is limited. There are physiological limitations. One can read a couple of papers published per day in one's special area (refer to arXive). How can one even read abstracts if hundreds of papers are being published in one's research topic? It will be difficult to keep oneself updated and unnecessary fragmentations (within sub-fields) occur.  Surely the cohesiveness of that field is under danger!

\item the phenomenon of ``re-discovering''. Literature survey takes time (lots of time when there are lots of papers to study). Easy way out: cut it! One remains ignorant about what has been published. Then the phenomenon of  ``rediscover it'' happens (most of the time unintentionally, but sometimes intentionally too, just to give a flavor of originality to one's own work)\cite{gian}.

\item fashionable/hot topics sometimes displace some important and core scientific problems. If some of the leading persons moves to a hot topic then majority of the community "jumps" to that hot topic, to ``catch-up'' in the race!  Is it good?  Will spirit of science gain from it? Many fundamental topics has been displaced by fashion! Dirac said: ``Quantum mechanics is a provisional theory.''  And many giants believed so. The problem of quantum mechanics has been displaced!  Practice of it is not in line with the market and fashion approach! The foundations of quantum mechanics is a difficult problem, working on it will not lead to quick and many publications.

There is another aspect to this issue. Consider a person very motivated and very deeply interested in the foundations of quantum mechanics. Will the current "market driven approach" support him? The answer is: "No"! The person will not get publications, and will be filtered out of the system.
Otherwise he/she has to do two types of research in parallel: One computational research which will lead to quicker publications and will support him/her in the system. And in the rest of the time he/she can work on the foundations of  quantum mechanics. But it requires lots of self discipline,  and very wise division of time and energy! Who is finally under loss? Foundations of quantum mechanics may be under loss!

\item loss of diversity; loss of free exploration; loss of independent minds; and loss of passion too!\cite{eva}.

\item unbearable pressure on editors and referees. Most researchers think that their work is excellent, and majority of them do all kinds of efforts and fights with the editorial board members to publish their results. This leads to loss of interest of editors and referees.

\item an introduction of ``selection staff'' (or professional editors) into editorial panels of reputed journals. As expert editors cannot do justice to hundreds of papers submitted per day, a new layer has been introduced in editorial boards of many journals.  This selection staff's duty is to reject! (sometimes 80-90 percent papers are rejected----arguing on the grounds of scope, and type of readership associated with the journal\cite{gian}).  Are the decisions of the selection staff always unbiased? No!  Who is responsible? overproduction!

\item plagiarism and clever plagiarism. There are uncountable cases of plagiarism. Overproduction and rush to publish has led to more and more cases of plagiarism. But thanks to recent powerful softwares that can trace out plagiarism! And hopefully, it is getting under control. However, "clever Plagiarism" is more "elusive". Stealing an already published idea and rewording it in a different form is a case of clever plagiarism\cite{gian}. Unfortunately, no software can catch it! Only moral values, and respect for your fellow researchers in the field can save your science from this evil.  Demand for a large number of publications is one of the reasons for these evil practices in science.

\een

\section{Phenomenon of predatory journals.}

Why are there so many predatory journals in recent times? Because there are so many papers produced that standard time-tested journals are not able to cope up with such floods of papers. They reject. They have to! This prompts paid and predatory journals, and this leads to the degradation of the quality and standards of scientific research.

\begin{figure}[!h]
\begin{center}
\includegraphics[height=6cm]{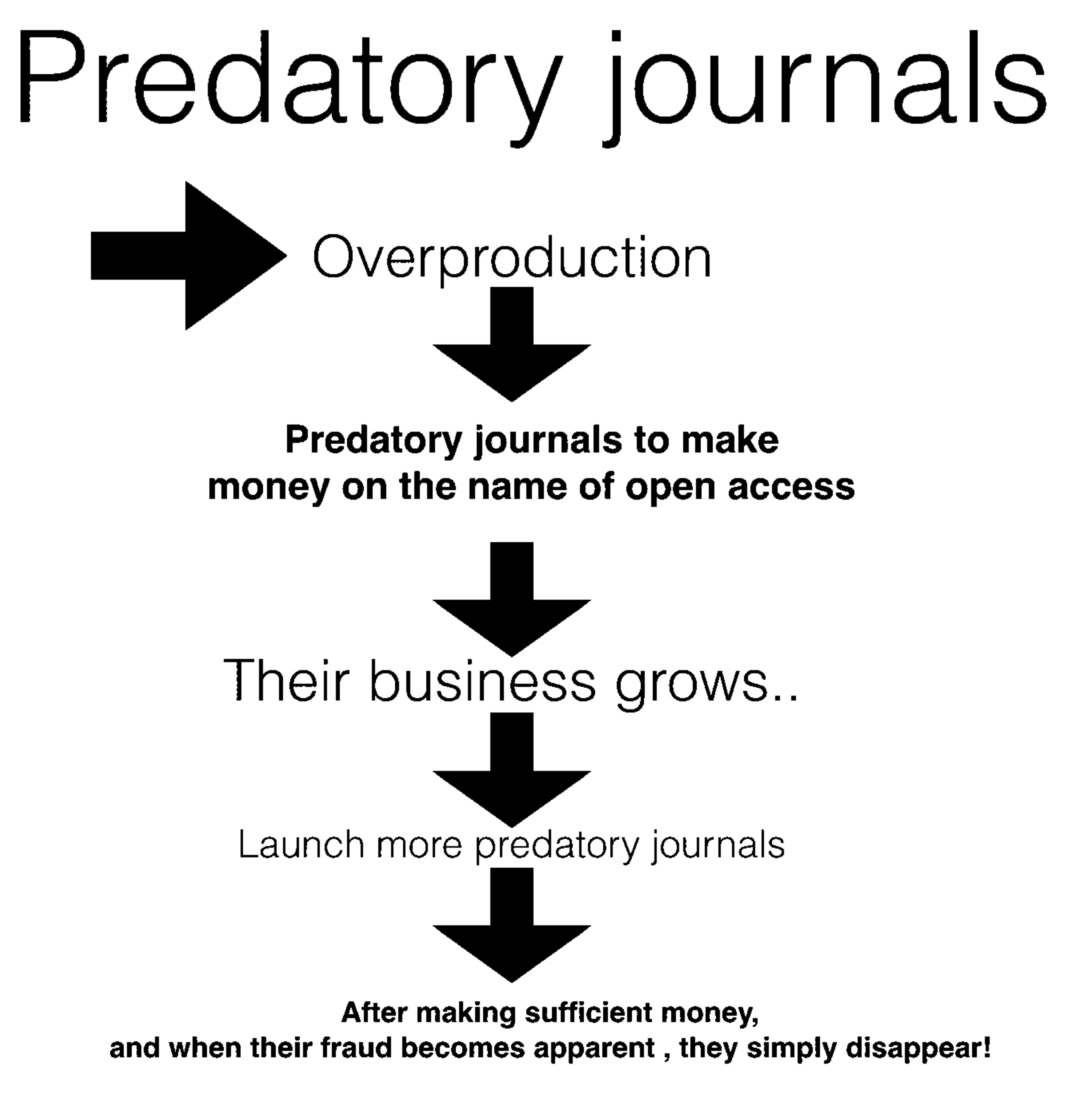}
\caption{Predatory journals and misuse of the open access.}
\end{center}
\end{figure}

Figure (4) depicts how the business of predatory journals grow. Overproduction is the root cause.

\section{Some Indices}

Surely the indices (impact factor of  a journal and h-index of a scientist) were created to improve the quality of scientific work and to improve scientific culture. But people always find loopholes and then these indices can be exploited and can be misused. Let us briefly go into these issues:

\subsection{Impact factor of a journal}
Impact factor (IF) of a journal is defined as: 
\beq
IF = \frac{Total~ number ~of ~citations~ for~ all ~the ~papers ~
published ~by~ a~ journal~ in ~a~ given~ year}{Total ~number~ of ~papers ~ published ~by ~that~ journal~ in~ that~ year}.
\eeq

It is a good parameter for a journal, but some commercial journals (whose main aim is to get profit) find ways to manipulate it. As Gianfranco\cite{gian} argues, editors of such journals invite review articles from respected/famous scientists. This leads to citations, and higher IF. In addition, editors tend to select papers that follow fashion, and have the potential to attract audience! This distorts the good scientific culture. This sometimes badly affects journals having more respect to moral values and good scientific culture. Those suffer. It is like one TV channel (which showcases entertainment and hot news) outperforming another TV channel (which is more culturally oriented (like classical stuff)).

So, only solution is that the scientific community should not be too obsessed by the impact factor business of scientific journals.

\subsection{h-index}

Can a single number define the research work of a scientist?  The answer is no! But in typical situations it is a good indicator. Problem comes when this is made the basis of promotions, getting jobs, giving funds, etc. Only a careful study of one's scientific work by an expert can evaluate that person's scientific output honestly. But it takes a lot of time! So, panels consisting of experts consider the easier counting method.  However,  it has obvious flaws. Just for example: h-index of Albert Einstein = 44. There are so many researchers whose h-index is >44! Can we compare? Peter Higgs has h-index=11. There are so many scientists whose h-index is greater than 11! Again, can we compare? No, we cannot compare this way. People who support h-index can argue that these are special cases, not general. Yes, they are right in that. But time tested good indicators: passion; novelty, and depth of works done; diversity; curiosity etc cannot be quantified and h-index misses these. These are the good indicators and are impossible to formulate in numbers!\cite{henri}.

\section{What is good in current times?}

The concept of arXive is the novel one! One gets immediate publicity and the ownership of ideas (the problems of the "theft of ideas" in the pre-arXive era is thus solved to some extent)\cite{henri}. Information technology and communications has made access to scientific publications very easy. Old culture of going to  the library, finding the hard copy of the required journal, finding the required article in it, and then getting it photocopied is all gone. With online access, one can read the soft copy in one's laptop or desktop, or one can get it printed on a printer. In fact, if you have good Internet connectivity, you can sit anywhere in the world and do theoretical research (You need some papers, a pencil, and the Internet, thats all!). The case of experimental research is different. The author would like to make a request to the community that if you find a good/appropriate arXive paper (whether it is published or not) please don't hesitate to cite it in your work.

\section{What should be done? Collective efforts?}

As mentioned before, data shows that the number of active scientists is growing fast (In 1930s: 1 scientist per 10,000 people. In 2013: 1 scientist per 700 people). But, the number of fundamental problems remains the same! Current scientific sociology resolves it by inventing and working on peripheral problems. This will not help science to advance further.

There should be some mechanism of knowing how many scientific positions/post doc positions etc are available (average, say, over a couple of years) and how many phd students are joining the academic research (again, average over a couple of years). This kind of data can balance the job opportunity equations,  and can lessen the pressure to compete. Thus can lessen the pressure to publish.

But it will be great if new riddles are discovered.  Real progress will happen when more scientists are interested in experimental and observational science! Because, it is the experimental and observational science that can discover new scientific problems/riddles worth working on! Scientific community has to seriously think about it! Currently, the division seems very skewed. Also, there has to be very tight correlation between the experimental work and the theoretical work.

The current academic culture of pressure and competition is not making students as true scholars. Current system is producing skilled manpower! Skilled to produce papers! If there is less pressure to publish, students can go to libraries and spend hours reading there, deepening and widening their knowledge of the topics of their interest. More stimulating and motivating environment is required.

How can the excessive load on some international journals be reduced? And how can the phenomenon of predatory journals be controlled. I think due respect to one's national journals must be given. Develop a culture where publishing in a national journal is a matter of prestige. If one of your national journal has got a very bad reputation (for publishing articles which have been rejected by almost all relevant journals), then time is ripe to launch a new journal! And forget the old one. National science academies must look into these matters, and should launch new journals (run by an academic staff), and set high standards of quality, and encourage researchers to send their manuscripts for publication. During jobs, publications in national journals should be given the weightage. It is like ``we love and respect our country''  then why don't  ``we love and respect our national journals?'' This will reduce excessive load on some international journals!

There should be more conferences organized by the academic staff (not by the commercial organizers) with plenty of discussion time. Be aware of predatory conferences (their sole purpose is to get financial profit)\cite{gian}.

\section{Conclusion}
I would like to end with a quote from Gianfranco Pacchioni\cite{gian}: ``The pressure towards achieving new results is daily, and leaves little room, if any at all, to ponder the meaning of what one does.''

Therefore, it is the time to publish less, novel and complete works. Only collective and world wide efforts can solve these grave problems of the practice  of science. Not you and me alone!

\section*{Acknowledgments}

This article is inspired by "Round table on science in h-index era", 22th--24th July, ICAM Global summit, 2020.



\end{document}